\documentclass[]{algotel}
\usepackage[latin1]{inputenc}
\usepackage[english]{babel}
\usepackage{amsmath}
\usepackage{txfonts}
\usepackage{graphicx}
\usepackage{graphics}
\usepackage{rotate}
\usepackage{float}
\usepackage{subfig}

\newtheorem{mydef}{Definition}
\newtheorem{mytheor}{Theorem}

\author{Hadrien Hours\addressmark{1}, Ernst Biersack\addressmark{1} \and Patrick Loiseau\addressmark{1}}

\title[Causal study of network performance]{Causal study of Network Performance}

\address{\addressmark{1}EURECOM, Campus SophiaTech, Les Templiers, 450 Route des Chappes, 06410 Biot Sophia Antipolis, France}

\keywords{Bayesian graph, Causality, Network Performance, TCP Protocol}

\begin{document}
\maketitle

\begin{abstract}
The use of Internet in the every day life has pushed its evolution in a very fast way. The heterogeneity of the equipments supporting its networks, as well as the different devices from which it can be accessed, have participated in increasing the complexity of understanding its global behavior and performance. In our study we propose a new method for studying the performance of TCP protocol based on causal graphs. Causal graphs offer models easy to interpret and use. They highlight the structural model of the system they represent and give us access to the causal dependences between the different parameters of the system. One of the major contribution of causal graphs is their ability to predict the effects of an intervention from observations made before this intervention.

\end{abstract}

\section{Introduction}
\label{sec:in}
%~\cite{McCreary2000}
The TCP protocol supports more than 80\% of the traffic going through the Internet and several studies of its behavior exist to better understand the different parameters and their roles in the performance one can expect when using this protocol. Theoretical models~\cite{Padhye:1998:MTT:285243.285291} as well as statistical ones~\cite{Mirza:2007:MLA:1254882.1254894} have been proposed. These models do not take into account the impact that parameters can have between each others and the approach is highly depending on the TCP version that is considered. For the second class of models, as relying on statistical correlation, there is the additional limitation of creating a model that is invalid as soon as it is used for predicting intervention that could change the statistics under which the model was inferred. The causal approach we propose in this paper answers these different issues.

The different works made in the domain of causal models inference and their representations as Bayesian graph, often Directed Acyclic Graphs (DAGs), as support for predicting interventions~\cite{PearlJ2000,citeulike:487589} give access to new perspectives in the study and understanding of complex systems.

While causality has been used to study network performance~\cite{Tariq:2008:AWD:1402946.1402971}, we place ourselves in a situation where resource, both in terms of data and computational power, is a limiting factor. We also introduce what we believe to be more formal and robust methods for both graph inference and intervention predictions.

In this paper we present the Causal Model Inference approach in Section~\ref{sec:introductioncausality} and show one example of its application to the study of FTP traffic in Section~\ref{sec:ftpstudy}. We will conclude this paper in Section~\ref{sec:conclusion}.

\section{Background}
\label{sec:introductioncausality}

Correlation is not causality, two parameters can be found correlated but this correlation does not give us any information on whether one is the cause of the other, or the opposite, or if there exists a third parameter \emph{causing} these two parameters. This basic notion illustrates the big difference between causal model and statistical model.

\subsection{Causal model inference}

In our work we use the PC algorithm~\cite{citeulike:487589} to infer our causal graph. This algorithm is based on leading independence tests to, gradually, build a graph where these independences are verified. The key, then, is the choice of the independence test that will assess the properties of our system. While the classical approach is to test partial correlation between the residuals of linear regression (as in Z-Fisher test), we are observing parameters which are not normally distributed and with non linear dependences. Consequently, in our work, we use the Hilbert Schmidt Independence Criterion~\cite{DBLP:journals/corr/abs-1202-3775}, a criterion that does not rely on normality or linearity to test parameters independences.

\subsection{Prediction}
\label{subsec:prediction}
Assuming now that we have our causal graph representing our system, it is possible, using simple graphical criteria~\cite{PearlJ2000,citeulike:487589}, to predict the effects of an intervention on one of the parameters of our model. 

In the graph each vertex corresponds to one parameter of our system and an edge between two vertices represents a direct causal effect from the parent to the child (in $X \rightarrow Y$, $X$ is the parent and $Y$ the child). A (set of) node(s), $Z$, is said to \emph{block} a path between $X$ and $Y$ if \textit{(i)} every collider ($\rightarrow W \leftarrow$), or its descendants, is not in $Z$ and \textit{(ii)} at least one non collider is in $Z$.

The notation $do(X=x)$ represents the manual intervention of setting the parameter $X$ to the value $x$.

\begin{mydef}[Back-door criterion]
 A set of variables, $Z$, is said to satisfy the Back-door criterion, relative to an ordered
pair of variables ($X_{i}$ ,$X_{j}$), in a DAG G if: \textit{(i)} No node in $Z$ is a descendant of $X_{i}$. \textit{(ii)} $Z$ blocks every path between $X_{i}$ and $X_{j}$ that contains an arrow into $X_{i}$
\end{mydef}

\begin{mytheor}[Back-door adjustment]
  If a set of variables $Z$ satisfies the Back-door criterion relative to ($X$,$Y$), then the
  causal effect of $X$ on $Y$ is identifiable and given by the formula:
  \begin{equation}
    P(Y=y\mbox{ }|{}\mbox{ }do(X=x)) = \sum_{Z} P(Y=y\mbox{ }|\mbox{ }X=x,Z=z)P(Z=z)
    \label{eq:backdooradjustment}
  \end{equation}
\end{mytheor}

Due to space constraints we will not present the theory of \emph{d-separation}, at the source of this criterion, but we redirect the reader to~\cite{PearlJ2000} for a formal explanation and justification of the Back-door criterion.

\section{Study of FTP traffic}
\label{sec:ftpstudy}
For illustrating our approach we decided to study FTP traffic. One reason for choosing this protocol is the absence of the application limiting the performance of TCP. The throughput is only limited by the Network or the client Receiver Window.

We first define the different parameters of our model and explain how do we build our dataset. Then, we present the causal model we obtain, with the PC algorithm, before predicting the effect of intervening on the \emph{RTT}, on the \emph{Throughput}.

\subsection{Dataset}
\begin{table}[t]
 \centering
 \small{
 \begin{tabular}{|c|c|c|c|c|c|}
    \hline
    \textbf{Parameter} & \textbf{Definition} & \textbf{Min} & \textbf{Max} & \textbf{Avg} & \textbf{Coeff. Var.}\\
    \hline
    Dist (km) & Distance between Server and Client & 14 & 620 & 250 & 0.95\\
    T.O.D. (s) & Time of the day, when the connection was started & 740 & 82000 & 46000 & 0.53\\
    Nbbytes (MB) & Number of bytes sent by the server & 6 & 60 & 31.5 & 0.51\\
    N.L.A.C. (kbps) & Narrow Link available capacity & 47.8 & 42.9e+3 & 5.9e+3 & 1.7\\
    Nbhops (units) & Number of hops between Client and Server & 9 & 27 & 11 & 0.24\\
    RTT (ms) & Round Trip Time & 60 & 710 & 270 & 0.76\\
    BufferingDelay (ms) & Part of the RTT due to queuing delay & 4.2 & 470 & 84 & 1.2\\
    RetrScore (no unit) & Fraction of retransmitted packets & 0.001 & 0.014 & 0.0018 & 0.92\\
    p (no unit) & Fraction of retransmitted windows of packets & 3e-5 & 0.011 & 0.0007 & 1.3\\
    T.O.R. (no unit) & Fraction of retransmitted packets due to Time Outs & 0 & 0.01 & 0.0006 & 1.7\\
    RWin (kbytes) & Receiver Window & 10.7 & 253.9 & 136.7 & 0.68\\
    Tput (kbps) & Throughput & 24 & 928 & 332 & 0.77\\
    \hline
 \end{tabular}}
  \caption{Summary of FTP traffic dataset}
  \label{tab:ds}
\end{table}

For our experiments we set up a FTP server where all the traffic is recorded. Using Intrabase~\cite{conf/e2emon/SiekkinenBUGP05} and the Tstat tool\footnote{http://tstat.tlc.polito.it/} we obtain the different metrics presented in Table~\ref{tab:ds}. This table also presents a summary of our dataset consisting in 1000 downloads from different clients, in Spain, Germany and France.

\subsection{Causal model}
\begin{figure}
 \centering
 \includegraphics[scale=0.4]{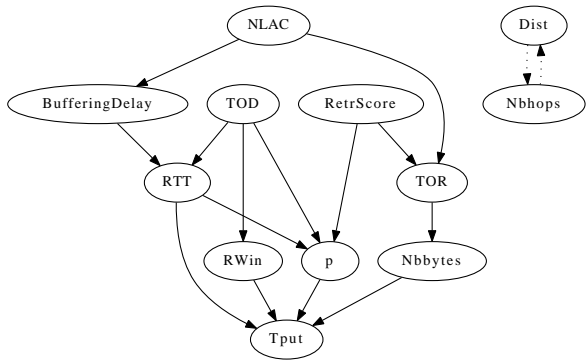}
 \caption{Causal graph model of FTP traffic performance}
 \label{fig:causalgraphftp}
\end{figure}

The model we obtained is presented in Figure~\ref{fig:causalgraphftp}. Due to space constraints we cannot describe all the properties of this model and focus on prediction, presented in the following section.

\subsection{Predictions}
    As we are working with variables where no distributions can be assumed, we use Copula Theory~\cite{jaworski2010copula} to estimate the densities, and conditional densities, of the parameters present in the Back Door adjustment formula, Equation (\ref{eq:backdooradjustment}).
    
    The presentation of Copulae, as well as the choice of the Copula to model the distributions of our parameters, are out of the scope of this paper.
    
   For predicting the effect of intervening on \emph{RTT}, we use the Back Door Criterion met by the set of variables $Z_{RTT}$, Equation (\ref{eq:bdrtt}), with $Z_{RTT}\mbox{ }=\mbox{ }\{T.O.D.,\mbox{ }N.L.A.C.\}$:
   \begin{eqnarray}
      f_{TPUT|do(RTT)}(Tput = \Delta | do(RTT=\theta)) = \int_{Z_{RTT}}f_{Tput|RTT,Z_{RTT}}(Tput = \Delta | RTT = \theta, Z_{RTT})f_{Z_{RTT}} \, \mathrm{d}Z_{RTT}
      \label{eq:bdrtt}
   \end{eqnarray}
    
\begin{figure}[ht!]
    \centering
    \begin{tabular}{rc}
    \rotatebox{90}{\hspace{1.4cm}$Throughput\mbox{ }(kbps)$} & \includegraphics[scale=0.2]{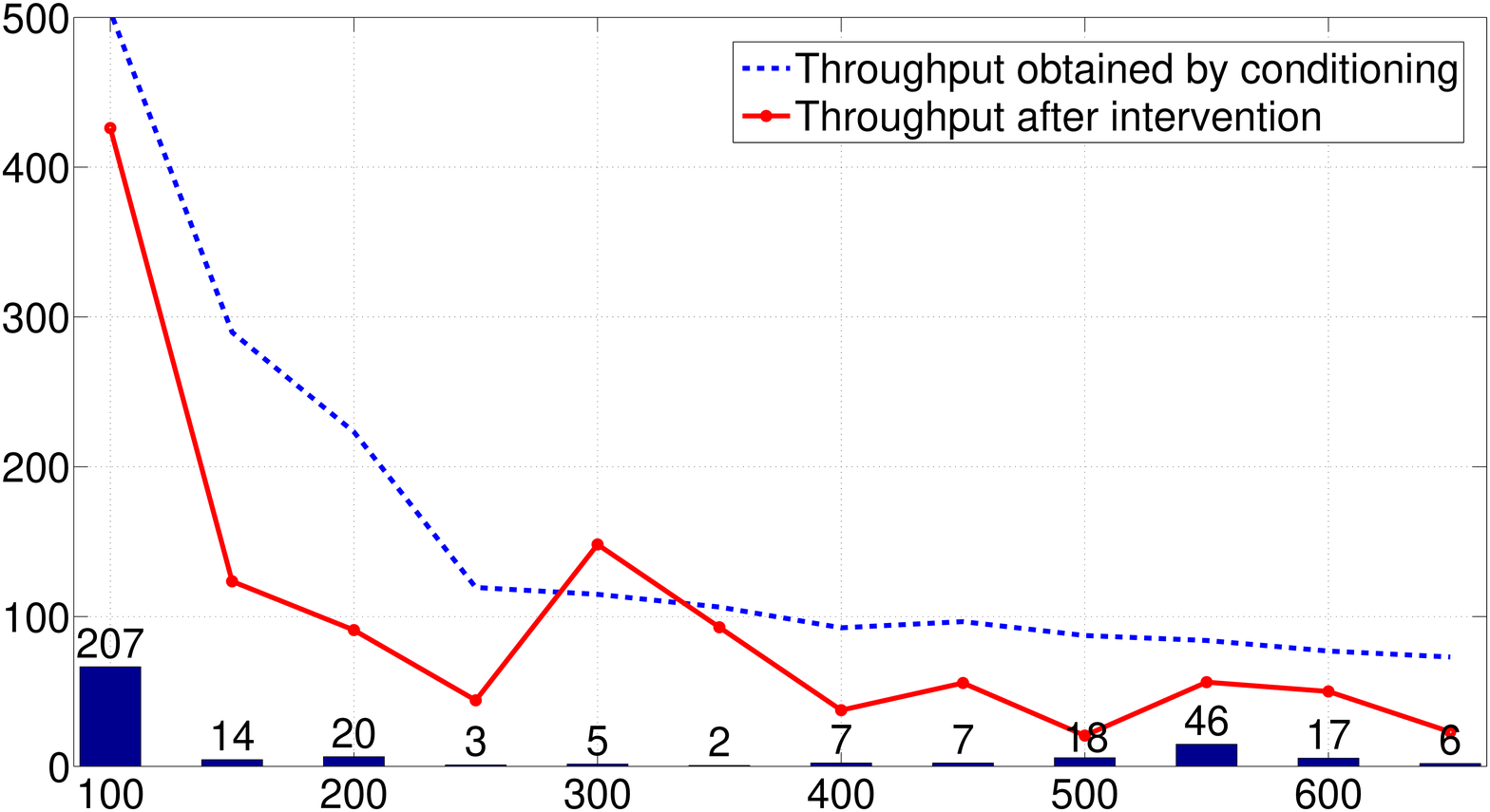}\\
     & $RTT\mbox{ }(ms)$
     \end{tabular}
    \caption{Comparison between causal approach and naive approach}
    \label{fig:tputcomparisons}
\end{figure}

   Figure~\ref{fig:tputcomparisons} presents the estimated \emph{Throughput} post-intervention (solid line), obtained with Equation (\ref{eq:bdrtt}). For comparison, we present the \emph{Throughput} obtained from observations, in the initial dataset, for which the \emph{RTT} value is the one corresponding to the intervention we want to predict (dotted line). We also added the information of the number of samples (bar plot) from which the distribution of the \emph{Throughput} post-intervention was estimated. For values of \emph{RTT} in $[250,350]$ we can see that there are not enough samples for estimating the post-intervention distribution which leads to inconsistent estimates.
   
   Removing the inconsistent estimates of the post-intervention distribution, we can see that the \emph{Throughput} post intervention is smaller than the one we observe in the initial dataset for a given value of \emph{RTT}. This observation can be explained by the causal graph presented Figure~\ref{fig:causalgraphftp} as, by conditioning on a given \emph{RTT} value, we also take into account the Back door effects of variables such as \emph{RWin}, \emph{N.L.A.C.}, \emph{T.O.D.} or \emph{T.O.R.} which are spurious associations blocked by $Z_{RTT}$ in Equation (\ref{eq:bdrtt}). This result shows that adopting a naive approach of estimating the \emph{Throughput} directly from the pre intervention samples will overestimate the effect that an intervention on the \emph{RTT} will have on the \emph{Throughput}.

\section{Conclusion}
\label{sec:conclusion}
This work is a first attempt to apply causal theory to the study of network performance. We present the inference of causal models, represented by Directed Acyclic Graphs, and their use to predict the effects of interventions on complex systems from passive observations. We show the different challenges that arise when applying theoretical theory to a real case study, the study of FTP traffic, and propose several solutions. As every model, causal models present their limits. We present the ones we judge as the most important for our work and propose solutions to overcome them. We introduce several methods that show very promising results. The use of the Hilbert Schmidt Criterion, for testing independences in the PC algorithm, and copulae, for estimating the multidimensional probability density functions, are the two main ones.

As we can see from Equation (\ref{eq:backdooradjustment}), the variety of situations observed defines the range of predictions that we can make. It will be necessary to increase the number of samples we have to reach a greater accuracy in our predictions. We are now working on network simulated experiments where we will have access to more parameters and the possibility to verify the accuracy of our predictions.

One important limitation of our approach is the definition of static parameters, by averaging metrics such as \emph{RTT} or \emph{RWin}, to model a dynamic protocol. Using the web10G tool, we plan to have access to low level TCP parameters and to sample the parameters at a finer grained timescale.

%\nocite{*}
\bibliographystyle{alpha.kris}
\bibliography{causalstudyofnetworkperformance.bib}
\label{sec:biblio}

\end{document}